\documentclass{elsart}

\usepackage{graphicx}
\usepackage{amssymb}
\usepackage{natbib}

\def\araa{{\em ARAA}}

\def\aj{{\em AJ}}

\def\apj{{\em ApJ}}
\def\apjs{{\em ApJS}}
\def\aap{{\em A\&A}}

\def\mnras{{\em MNRAS}}

\def\prl{{\em Phys.\ Rev.\ Lett.}}

\def\pasp{{\em PASP}}

\def\rmp{{\em Rev.\ Mod.\ Phys.}}

\def\astrobj#1{#1}

\begin{document}

\begin{frontmatter}

\title {MODEST-2: A Summary}

\author{Alison Sills},\address{Department of Physics and Astronomy, McMaster University, Hamilton, ON L8S 4M1, Canada; asills@physics.mcmaster.ca}

\author{Stefan Deiters},\address{Astronmisches Rechen-Institut, M{\"o}nchhofstrasse 12-14, D-69120 Heidelberg, Germany }

\author{Peter Eggleton}, \address{L-413, Lawrence Livermore National Laboratory, 7000 East Avenue, Livermore, CA 94551 USA}

\author{Marc Freitag},\address{Astronmisches Rechen-Institut, M{\"o}nchhofstrasse 12-14, D-69120 Heidelberg, Germany }

\author{Mirek Giersz},\address{Nicolaus Copernicus Astronomical Center, Polish Academy of Sciences, 00-716 Warsaw, ul. Bartycka 18, Poland}

\author{Douglas Heggie},\address{University of Edinburgh, School of Mathematics, King's Buildings, Edinburgh EH9 3JZ, UK}

\author{Jarrod Hurley}, \address{Department of Astrophysics, American Museum of Natural History, Central Park West at 79th Street, New York, NY 10024, USA}

\author{Piet Hut}, \address{Institute for Advanced Study, Princeton, NJ 08540, USA}

\author{Natasha Ivanova}, \address{Northwestern University, Department of Physics and Astronomy, 2145 Sheridan Rd. Evanston, IL 60208 USA}

\author{Ralf S. Klessen}, \address{Astrophysikalisches Institut Potsdam, An der Sternwarte 16, D-14482 Potsdam, Germany}

\author{Pavel Kroupa}, \address{Institut fuer Theoretische Physik und Astrophysik, Universitaet Kiel, D-24098 Kiel, Germany}

\author{James C. Lombardi Jr.}, \address{Department of Physics and Astronomy, Vassar College, 124 Raymond Avenue, Poughkeepsie, NY, 12604, USA}

\author{Steve McMillan}, \address{Department of Physics,3141 Chestnut St., Drexel University, Philadelphia, PA, 19104, USA}

\author{Simon Portegies Zwart},\address{Sterrenkundig Instituut ``Anton Pannekoek", Kruislaan 403, 1098 SJ Amsterdam, the Netherlands}

\author{Hans Zinnecker}\address{Astrophysikalisches Institut Potsdam, An der Sternwarte 16, D-14482 Potsdam, Germany}

\begin{abstract}
This is a summary paper of MODEST-2, a workshop held at the 
Astronomical Institute ``Anton Pannekoek'' in Amsterdam, 16-17
December 2002. MODEST is a loose collaboration of people interested
in MOdelling DEnse STellar systems, particularly those interested in
modelling these systems using all the available physics (stellar
dynamics, stellar evolution, hydrodynamics and the interplay between
the three) by defining interfaces between different codes. In this
paper, we summarize 1) the main advances in this endeavour since
MODEST-1; 2) the main science goals which can be and should be
addressed by these types of simulations; and 3) the most pressing
theoretical and modelling advances that we identified.
\end{abstract}

\end{frontmatter}

\section{Introduction}

Dense stellar systems can roughly be defined as environments in
which the interactions between individual stars play a crucial role.
At the very least, two-body relaxation is short enough to have changed
the stellar distribution function significantly since the formation of
the system; and in the more interesting cases, actual stellar
collisions have changed the properties of individual stars. These two
effects are related: relaxation can lead to the dynamical segregation
of binaries in the core, increasing the rate of encounters and the
temporary capture of single stars or members of other binary stars,
providing episodes that enhance the probability of physical collisions
between stars.

Our observational insight into dense stellar systems has advanced
enormously during the last decade.  The Hubble and Chandra telescopes
have allowed us to peer into the dynamical heart of the densest
globular clusters, we have found stars orbiting the central black hole
in our galaxy, and infrared observations have penetrated into the most
obscured areas of star forming regions, where protostars can
physically interact with each other before settling down as relatively
more isolated stars.

From a theoretical point of view, the challenge has long been to model
the ecological network of interactions coupling the stellar evolution and
stellar dynamics of dense stellar systems.  While the 1980s
saw more and more detailed dynamical models of globular clusters, it
was only in the later '90s that these models started to
incorporate some approximate form of stellar evolution, based on
fitting formulae, and binary star evolution, based on recipes.  The
next step will be to model the merging process of colliding stars more
accurately, and to incorporate more detailed forms of on-the-fly
modelling of the stellar evolution of the dynamical merger products.

Simulations of dense stellar systems currently face two major hurdles,
one astrophysical and one computational.  The astrophysical problem
lies in the fact that several major stages in binary evolution, such
as common envelope evolution, are still poorly understood.  The best we
can do in these cases is to parameterize our ignorance, in a way that
is reminiscent of the introduction of a mixing length to describe
convection in a single star, or an alpha parameter in modelling an
accretion disk.  The hope is that by modelling a whole star cluster in
great detail, and comparing the results to the wealth of observational
data currently available, we will be able to constrain the parameters
that capture the unknown physics.

There is an interesting parallel with the way single stars were
modeled, notwithstanding the fact that there were uncertainties in
various aspects of its microscopic physics.  An early triumph of
stellar evolution was the prediction of an excited state in the
${}^{12}$C nucleus, in order to reconcile the results of stellar
evolution calculations with observations, a prediction that soon
afterward was confirmed in the laboratory.  And more recently,
neutrino mixing has been confirmed as the explanation for a
long-standing discrepancy between the standard model for the evolution
of the \astrobj{Sun} and direct observations of the neutrino flux
coming from the core of the \astrobj{Sun}.

Our hope is that the more complex modelling of whole star clusters
will similarly shed light on the `microphysics' input, in this case
the poorly known fate of complex stages of binary star evolution.  But
in order to constrain scenarios for common envelope evolution, for the
formation of millisecond pulsar binaries, etc., we need to construct
accurate models for the evolution of dense stellar systems.  This
brings us to the second major hurdle, which is of a computational
nature.  The problem is one of composition: while we have accurate
computer codes for modelling stellar dynamics, stellar hydrodynamics,
and stellar evolution, we currently have no good way to put all this
knowledge together in a single software environment.

It was the goal of the MODEST-1 meeting, in New York City in the
summer of 2002, to begin addressing this problem.  The MODEST acronym
was coined during this meeting, and it can stand not only for MOdelling
DEnse STellar systems, but also for MODifying Existing STellar codes.
The latter description stresses the desirability to start with what is
already available, and to find ways to put it all together, rather
than to try to write a kitchen-sink type over-arching super code from
scratch.  We refer to the MODEST-1 review paper for further background
\citep{MODEST1}, and also to the MODEST website: {\tt
http://www.manybody.org/modest.html}.  The present paper offers a
summary of the presentations and discussions held during the MODEST-2
workshop, organized at the University of Amsterdam, Holland, by Simon
Portegies Zwart and Piet Hut, in December 2002.  This paper contains
the input of many participants, who are listed below under the
acknowledgments.  While many of the authors have contributed to
various sections, each section has main author(s), as follows.  \S1
was written by Piet Hut, \S2 was written by Marc Freitag, Mirek
Giersz, Stefan Deiters, Natasha Ivanova, James Lombardi \& Steve
McMillan, \S3 was written by Ralf Klessen, Pavel Kroupa \& Hans
Zinnecker, \S4 was written by Steve McMillan, Jarrod Hurley, Peter
Eggleton, Simon Portegies Zwart \& Alison Sills, \S5 was written by
Douglas Heggie, and \S6-7 were written by Alison Sills.

MODEST-2 was an informal workshop, consisting of 8 short talks from
participants outlining how their work fits into the MODEST framework,
what they want to get out of participation in MODEST, what the most
relevant questions are for their area, or what they have accomplished
since MODEST-1. There was also a fair amount of time allocated for
general discussion of science goals and short-term theoretical goals
before the next MODEST meeting. Finally, we spent some time discussing
the long-term goals and best way to future of the collaboration.

The biggest difference between MODEST-1 and MODEST-2 was a
concentration on WHAT rather than HOW. MODEST-1 was spent deciding
that the different communities (evolution, dynamics, hydrodynamics)
could and should work together, and then discussing exactly how they
wanted to do that -- the details of the interface. At MODEST-2, we
spent some time discussing the interfaces and their implementation
(see \S2.4) but most of the time was spent talking about the
different scientific issues that the MODEST collaborators wanted to
see addressed. This paper attempts to capture the tone of the
meeting, and outline the current state of MODEST research.

\section{Progress since MODEST-1}

MODEST-2 was held six months after MODEST-1. In that time, some
advances were made on combining stellar evolution, stellar dynamics
and hydrodynamics in modelling dense star clusters. In addition, some
groups have made progress in other areas of cluster modelling that
are relevant to the MODEST collaboration. In this section we summarize
the recent work on both these fronts. In \S2.1, Marc Freitag and
Natasha Ivanova discuss Monte Carlo codes that include the effects
of stellar collisions. In \S2.2, Mirek Giersz discuses the
alternatives of using scattering cross sections and ``live'' few-body
integrations in hybrid codes. Stefan Deiters describes the gaseous
codes in \S2.3. And in \S2.4, Jamie Lombardi, Steve McMillan and
Jarrod Hurley discuss their implementation of the MODEST-1 interface
between stellar dynamics, hydrodynamics and stellar evolution.

\subsection{Monte Carlo cluster simulations with stellar collisions }

\subsubsection{A Monte Carlo code for galactic nuclei simulations}

In the past few years, a new Monte Carlo (MC) code has been developed
to follow the long term evolution of galactic nuclei
\citep{FB01a,FB02b,Freitag01}. This tool is based on the scheme first
proposed by \citet{Henon73} to simulate globular clusters but, in
addition to relaxation, it also includes collisions, tidal disruptions
by a central massive black hole (BH), stellar evolution and captures
of stars by a central BH through emission of gravitational waves.

The MC technique assumes that the cluster is spherically symmetric and
represents it as a set of particles, each of which may be considered
as a homogeneous spherical shell of stars sharing the same orbital and
stellar properties. The number of particles may be lower than the
number of stars in the simulated cluster but the number of stars per
particle has to be the same for each particle.  Another important
assumption is that the system is always in dynamical equilibrium so
that orbital time scales need not be resolved and the natural
time-step is a fraction of the relaxation (or collision) time. 
The relaxation is treated as a diffusive process \citep{BT87}. 

Contrary to methods based on an integration of the Fokker-Planck (FP)
equation, with which it shares most assumptions, the particle-based MC
approach allows for a more direct inclusion of further physics, like
collisions, tidal disruptions, captures, large-angle scatterings or
interaction with binaries.  Other advantages over the FP codes include
the fact that the MC scheme handles a continuous stellar mass spectrum
and an arbitrary (anisotropic) velocity distribution without added
difficulty. Thank to a binary tree structure that allows quick
determination and update of the potential created by the particles,
the self gravity of the stellar cluster is accurately accounted for. 

The CPU time required by direct $N$-body simulations scales with the
number of particles $N$ like $N^{2-3}$, thus imposing a limit on $N$
of order a few $100\,000$, even with special-purpose GRAPE
computers. In contrast to this, MC runs, whose CPU time scales like
$N\ln(N)$, routinely use 500\,000 to a few millions of particles on
run-of-the-mill PCs. Such high numbers of particles mean that, for the
first time, globular clusters can actually be modelled on a star by
star basis \citep{Giersz98,Giersz01,JRPZ00,JNR01,WJR00}.

\subsubsection{Including stellar collisions}
\label{sec:coll}

Collisions between main sequence (MS) stars are treated with a high
degree of realism through the use of a comprehensive set of $\sim
15\,000$ SPH
\cite[Smoothed Particle Hydrodynamics,][]{Benz90} simulations
\citep{MaThese,FB00a,FB03}. Reducing this huge amount of data into a set of 
fitting formulae giving the outcome of a stellar solution as a
function of its initial conditions (the masses of the stars, the
relative velocity and the impact parameter) has so far proven
inconclusive. Thus, an interpolation scheme was used, based on a
Delaunay tessellation of the 4D, irregularly populated initial
parameter space to produce a 4-index lookup table. Interestingly, it
appears that the collisional mass loss, as determined by SPH simulations
is quite precisely predicted by a very simple semi-analytical model
of collisions, first proposed by \citet{SS66}, that considers only
conservation of momentum and total energy, as soon as the relative
velocity at infinity is higher than the escape velocity from the
surface of a star and the impact parameter is larger than about
$0.5(R_1+R_2)$ where $R_{1,2}$ are the stellar radii. This regime is
mostly relevant to collisions in a galactic nucleus, near the central
BH. This raises hope that some quick semi-analytical way of
treating high-velocity collisions can be devised that would complement
the work done Lombardi and his collaborators for parabolic encounters
\citep{LWRSW02}. Of particular interest would be the development of some
entropy-sorting algorithm to determine the post-collisional stellar
structure. This information is indeed required to compute the
subsequent evolution of stars that have suffered from
collisions. Unfortunately, it is doubtful that it can be extracted
from Freitag's SPH simulations that are of relatively low resolution and
make use of unequal mass particles, two facts that may lead to
important spurious mixing according to
\citet{LSRS99}. 

As shown independently by \citet{1991PhDT........11R} and
\citet{1993ApJ...404..717H}, the usual formulations of SPH that use
variable smoothing lengths fail to conserve energy and entropy
simultaneously.  However, \citet{2002MNRAS.333..649S} have recently
derived SPH equations of motion that, by construction, conserve both
energy and entropy even when the smoothing is adaptive.  The
derivation utilizes a Lagrangian, with Lagrange undetermined
multipliers employed to satisfy the constraint that the total mass
within the smoothing volume of each particle be held constant.
Although the corrections introduced by this new method become
vanishingly small as the number of particles $N\rightarrow\infty$, it
does seem to be a fundamentally better formulation of SPH.  Live (that
is, on-the-fly) SPH calculations in a cluster simulation, for example,
could benefit significantly from such a method, as they could achieve
higher accuracy for a fixed (and presumably relatively small) number
of particles.

In the simulations of \citet{MaThese}, either of two very simple
assumptions were used to set the stellar evolution of mergers. {\em
(1) Complete rejuvenation.} The merger is assumed to be completely
mixed during the collision and is put back on the zero-age MS. This is
quite unphysical and obviously leads to an important overestimate of
the merger's MS life-time but corresponds to the assumption made in
many previous works \cite[][for instance]{QS90}. {\em (2) Minimal
rejuvenation.} In this case, during a coalescence, the
helium cores of both parent stars merge together, while the hydrogen
envelopes combine to form the new envelope; no hydrogen is brought to
the core. An effective age is assigned to the merger by using a linear
relation for the mass of the helium core as a function of the time
spent on the MS and resorting to ``normal'' stellar evolution models
to provide the mass of the helium core at the end of the MS
\citep{HPT00,BelczynskiEtAl02}. In both cases, if the stars don't
merge no rejuvenation is assumed. Also, the thermal time scale is
always assumed to be much shorter than the average time between
collisions so that the MS mass--radius relation is applied to
collisions products.

\subsubsection{A route to intermediate mass black holes}

Many scenarios have been proposed for the formation of massive BHs in
the center of dense stellar clusters \citep{BR78}; most of them
require further investigation. Here, we explore the growth of a very
massive MS star (a few $\times 10^2$ to $\sim 10^4\,M_\odot$) by
run-away merging of stars \citep{RF03}. If its metallicity is
sufficiently low, such an object is likely to form an intermediate
mass BH (IMBH, with $M_\mathrm{BH}\simeq 100-10^4\,M_\odot$) at the
end of its life \citep{FK01,WHW02}.  This run-away route has been
shown to operate in FP models of simple proto-galactic nucleus models
by \citet[][hereafter QS90]{QS90} and in $N$-body simulations of
populous young clusters by \citet{PZMcM02}. In the later case, stellar
collisions occur in dynamically formed binaries and the authors argue
that the condition for run-away to occur is that the time scale for
the most massive stars ($M_\ast\simeq 100\,M_\odot$) to segregate to
the center of the cluster, $T_\mathrm{segr}$, be shorter than their MS
life-time, of order 3\,Myrs. Freitag's MC code cannot account for
binaries. This is not a serious concern because their formation and
survival in high-velocity galactic nuclei is unlikely. As the stellar
density rises to higher and higher values during the
(segregation-driven) core collapse, collisions are bound to occur even
without the mediation of binaries.

For definiteness, we concentrate here on QS90's model E4A, a Plummer
cluster with initial central values of the density and of the 3D
velocity dispersion of $3\times 10^8\,M_\odot\mathrm{pc}^{-3}$ and
$400\,\mathrm{km}\,\mathrm{s}^{-1}$. QS90 started their FP simulations
with all stars having $1\,M_\odot$ and assumed that all collisions
lead to mergers and that complete rejuvenation is valid.  Not
surprisingly, if we use the same, unrealistic, treatment of collisions
as QS90, we get clear run-away growth of one or a few particles. When
we switch to the realistic SPH prescription for the collisions and
minimal rejuvenation, we still get run-away. However, if we initiate
the cluster with a more realistic extended IMF \citep{Kroupa00b},
important mass loss from the massive stars occurs before core collapse
has proceeded to high stellar densities. As we assume that the gas is
not retained in the cluster, this mass loss drives the re-expansion of
the whole system. A second, deeper core collapse occurs later, when
the stellar black holes segregate to the center. The subsequent
evolution of this dense cluster of stellar BHs cannot be treated with
Freitag's MC code because dynamically formed binaries will play a
central role. Whether an IMBH may grow in such an environment is a
debated issue \citep{PZMcM00,MH02}.

In addition to models with the same densities and velocity dispersions
as considered by QS90, Freitag also simulated clusters with densities 3 and
9 times larger with correspondingly shorter relaxation times and, hence
$T_\mathrm{segr}$. Run-away growth happens in all simulations with
$T_\mathrm{segr}<3\,\mathrm{Myrs}$ but in none of the other cases. The
growth of the run-away particle(s) is limited to a few $100\,M_\odot$
($650\,M_\odot$ in the ``best'' case), probably by some still
unelucidated numerical artifact.  Note that 500\,000 particles were
used for these computations, independently of the number of stars to
simulate. Hence, every particle represents many stars (12 to 36 for
the simulations discussed here), a numerical treatment whose validity
becomes obviously questionable as soon as a single particle detaches
from the overall mass spectrum. Anyway, before the
run-away particle abruptly stops accumulating mass, its growth is
extremely steep. Once started, it occurs on a time scale much shorter
than stellar evolution and it seems that it can only be terminated by
some instability setting in in the structure of the massive star, the
inefficiency of collisional merging\footnote{Freitag hasn't computed SPH
collision simulations for stars more massive than $75\,M_\odot$ so
considerable extrapolation of the results is required.}, the depletion
of the ``loss-cone'' orbits that bring stars to the center or some
combination of these factors. Despite these uncertainties, stating that run-away merging produces stars at least as massive
as $500\,M_\odot$ in the center of clusters with
$T_\mathrm{segr}<3\,\mathrm{Myrs}$ is a robust conclusion.

\subsubsection{Monte Carlo codes for Globular Cluster Evolution}

The Monte Carlo code {\tt StarFokker}, being developed by A. G\"urkan
and F. Rasio at Northwestern and J. Fregeau at MIT (see Joshi et
al. 2000, 2001; Fregeau et al. 2002; Waters et al. 2000), currently has
the following features: fast integration of large numbers of stars (up
to $4\times 10^6$ stars for a Hubble time in about a week of computing
time), tidal truncation of the cluster, simple treatment of stellar
collisions (sticky sphere approximation), binary-binary interactions
with simple recipes (based on the previous Fokker-Planck study by Gao
et al. 1991), binary-single interactions with direct integration
(using {\tt scatter3} from STARLAB) and single star evolution (based
on Hurley et al. 2001). Work is in progress to incorporate a new
4-body integrator (developed by J.  Fregeau) for binary-binary
interactions, as well as a full treatment of binary star evolution
based on the population synthesis code {\tt StarTrack} (developed by
K. Belczynski and V. Kalogera at Northwestern; Belczynski et
al. 2002).  A new study of equal-mass clusters with primordial
binaries was recently completed (Fregeau et al. 2003), showing that,
in an isolated cluster, primordial binary burning can easily support
the cluster against core collapse for many Hubble times as long as the
initial binary fraction is larger than a few percent. After the
initial core collapse, gravothermal oscillations powered by the
remaining primordial binaries are always observed. The Monte Carlo
simulations also show the temperature inversion in the core expected
during re-expansion (Makino 1996, Giersz 1998).  In tidally truncated clusters with
primordial binaries, the models suggest that complete disruption of
the cluster often happens before core collapse.  Comparisons between
the simple recipes and direct dynamical integrations for 3-body
(binary-single) interactions show that the recipes are reasonably
accurate. However, binary-binary interactions are dominant for the
evolution of most cluster models with initial binary fractions above a
few percent.

A new Monte Carlo code, IMGE, that incorporates a lot of the new
ideas discussed at the MODEST-1 workshop, is being developed by A. G\"urkan
at Northwestern. Initial conditions are handled as in STARLAB, and the code
uses the FITS format to store snapshots that can be read back in.
Currently this code can only treat the evolution of an isolated cluster of
single stars.

\subsection{Hybrid Code -- Cross Sections for three- and four-body interactions}

Spherically symmetric equal mass star clusters containing a
large amount of primordial binaries are studied using a hybrid method,
consisting of a gas dynamical model for single stars and a Monte Carlo
treatment for the relaxation of binaries and three- and four-body
encounters. The initial conditions are as follows: a cluster of 300 000
single stars and 30 000 binaries, both distributed in Plummer's model
density distribution with a constant density ratio between binaries and
single stars. All binaries are set up with a so-called thermal
eccentricity distribution, and binding energies are equally logarithmically
distributed between 3 and 400 KT. Each binary-single star/binary
encounter is investigated by means of a highly accurate direct few-body
integrator (kindly supplied by S.J.~Aarseth with his  NBODY6 program
package). Hence hybrid codes can study the systematic evolution of individual
binary orbital parameters and differential and total cross sections for
hardening, dissolution or merging of binaries from a sampling of
several ten thousands of scattering events as they occur in real
cluster evolution (see Giersz \& Spurzem 2003 for details).

For three-body encounters Giersz \& Spurzem find a good agreement of
the nearly entire differential cross section with Spitzer's (1987)
expression, except for very small energy changes. This is not
surprising, because of the limited coverage of phase space for all
encounters with small energy changes in real cluster models compared
to artificial experiments. The formation of bound three-body
subsystems and binary dissolutions are not very probable. Merging
(interactions with minimum distance smaller than 1 $R_{\odot}$), as
expected, occurs preferentially at high $\Delta$ (relative binary
binding energy change). For smaller $\Delta$ non-merging encounters
dominate.

For four-body encounters, the hybrid code results are in good
agreement with Spitzer's (1987) and Heggie's (1975) analytical
formulae for $\Delta > 0.1$. For smaller $\Delta$, as it was predicted
by Heggie's (1975) analytical work for a tidal, adiabatic encounter
the differential cross section is proportional to
$1/(\Delta)log(\Delta)^{1/3})$. For strong encounters hardening of one
binary and dissociation of another dominates. For $\Delta < 1$ there is
a competition between dissociation and stable end configurations
(resulting in two surviving binaries). At small energy changes
formation of bound quadruples and stable hierarchical triples is the
most probable reaction channel. It is interesting to note that
Spitzer's and Heggie's formulae for three-body interactions also
describe with good accuracy four-body interactions.

For the first time, our study gives a complete overview of the behavior
of eccentricities in binaries embedded in an evolving star cluster. We
also find a new approximate law to fit our empirical cross sections for
eccentricity changes. The effects of flybys and close encounters can be
clearly distinguished. For the three-body encounters, for initially
nearly circular orbits, all final eccentricities after a
three-body encounter occur with equal probability. If there is already
some initial eccentricity the probability to reach any higher
eccentricity is approximately constant, while the chance to go back to
a less eccentric orbit decays exponentially ($\propto \exp(4 e_{\rm
init})$). This is even more pronounced for initially highly eccentric
binaries. For the four-body encounters a bimodal distribution of final
eccentricities, depending on whether we look at strong encounters or at
weak ones, can be seen. For strong encounters, the
initial eccentricity is ``forgotten'' in the sense that all
differential cross sections have a maximum at high final eccentricities
and decay again with the characteristic law seen already in three-body
encounters. For weak encounters (fly-bys) there is no strong
interaction and hence no strong eccentricity change. Finally, it is
interesting to note that in all evolutionary stages a so-called
thermal eccentricity distribution is maintained at all binary binding
energies.

\subsection{Gaseous Models}

The gap between direct models and the most interesting particle
numbers in real globular star clusters can until now only be bridged
by theory. The gaseous model (Louis \& Spurzem 1991,
http://www.gaseous-model.de) for example makes use of the remarkable
resemblance between a star cluster containing a large number of stars
and a self-gravitating gaseous sphere with a huge number of atoms (a
generalization to axisymmetric systems has not yet been tackled). Its
model equations are obtained as a set of moment equations of the local
Fokker-Planck equation. Compared to direct solutions of the
orbit-averaged Fokker-Planck equations it is easier to add new physics
and faster standard numerical solvers can be used (see for a
comparison e.g. Giersz \& Spurzem 1994).

The gaseous model played an important role in theory, but up to now it
has not been used to model observations directly. Concepts of
gravothermal contraction and oscillations (Lynden-Bell \& Eggleton
1980, Bettwieser \& Sugimoto 1984) have been derived in the context of
gaseous models and have proven to be very useful even now in the time
of huge direct $N$-body modelling. Comparisons between the different
models have produced promising results, so the time has come to
improve the gaseous model in order to get a more realistic model that
is capable of modelling real star cluster observations. In a first step
the effects of stellar evolution in the model were included using the
stellar evolution routines of Hurley, Pols \& Tout (2000) and
generated artificial color-magnitude diagrams. Although these diagrams
cannot be compared with observed ones, one gets a first idea of the
strength of the model: For example one can observe how population
gradients develop (heavy remnants sink to the center and low mass
stars migrate to the outskirts). More features need to be included,
among them kicks of neutron stars, a tidal field, dynamically active
binaries, collisional cross sections and binary star evolution in the
code. This would make the gaseous model a powerful tool to model
observations of globular star clusters. It could be also used to
conduct huge parameter studies in order to find a set of initial
parameters for higher precision models (Deiters, Hurley, \& Spurzem,
2003).

\subsection{The Stellar Dynamics -- Stellar Evolution -- Hydrodynamics Interface}

One of the goals of the MODEST-1 workshop was to specify ways to let
existing computer codes for stellar dynamics (SD), stellar evolution
(SE) and stellar hydrodynamics (SH) communicate with minimum
modification. With a well-defined minimal interface, each of the three
modules should see the others as a black box.  For example, the SD
module should not care whether the SE data result from running a live
SE code, or from a look-up table or fitting formula.

Immediately following the first workshop, Hut and Makino wrote a toy
model version for the SD-SE interface.  In order to test their
interface, they constructed a very simple implementation of both the
SD and SE parts of a simulation code.  For the SD they envisioned two
unbound stars on a head-on collision course that merge into a single
star with an unusual composition.  If mass loss during the collision
is neglected, and if the collision product is approximated as fully
mixed, then the SH module is effectively bypassed.  Their SE code then
approximates the stellar mass, radius and chemical compositions of the
collision product with a piece-wise linear function in time, with one
discontinuity.  A more detailed description of their SE module, as
well as the source code in both Fortran and C++, is publicly available
online\footnote{\tt http://www.manybody.org/modest\_star.html}.

The intent of Hut and Makino is that their code would be the
instigator of an ongoing effort in which the physics within each
module will be improved upon by the experts in that field.  The first
incremental refinement made was to include a non-trivial treatment of
the SH: the resulting program, dubbed TRIPTYCH\footnote{\tt
http://faculty.vassar.edu/lombardi/triptych}, uses the Make Me A Star
(MMAS) software package to determine mass loss during collisions as
well as the structure and composition profiles of collision products.
MMAS implements fast fluid-sorting algorithms to treat nearly
parabolic encounters between low-mass main sequence stars
\citep{LWRSW02}.  The source code for both MMAS and TRIPTYCH is freely
available from their web sites.

In order to improve the SE in TRIPTYCH, Hurley wrote wrappers to his
single-star evolution (SSE) code \citep{HPT00}, closely following the
SE interface defined by Hut and Makino.  The SSE routines use analytic
fitting formulae to approximate accurately the evolution for a broad
range of stellar masses and metallicity.  For the SH to interface with these SE
routines, it is still necessary to assume that the product becomes
fully mixed immediately after the collision, an assumption that cannot
be relaxed until a live SE code is introduced.

TRIPTYCH can be run online via a web interface, 
originally developed by Vicki Johnson of Interconnect.  The user
simply chooses two stellar models from a drag-down list, and enters
values for a relative velocity, periastron separation and initial
separation of the parent stars.  Within just a few seconds, the output
of TRIPTYCH is displayed, including plots of the orbital dynamics of
the parent stars, as well as the stellar profiles and the subsequent
evolutionary track on an HR diagram for the collision product.

An outgrowth of TRIPTYCH is a sister program, called TRIPLETYCH\footnote{\tt http://faculty.vassar.edu/lombardi/tripletych}, that
simulates the interaction of three stars, including the orbital
trajectories, possible merger(s), and the subsequent evolution of the
merger product.  TRIPLETYCH is one star closer than its counterpart
TRIPTYCH toward a realistic simulation of a star cluster.  McMillan
has implemented the SD of the three parent stars in TRIPLETYCH using
the scatter3 routine from STARLAB, with visualizations generated by
the snap\_to\_image routine.  Two of the stars are initially bound,
with the third approaching from infinity.  The scattering package is
described in detail by McMillan \& Hut (1996).  All
orbital parameters may be specified by the user; those left unset are
chosen randomly from appropriate distributions.

All STARLAB scattering packages (scatter3 and its higher-order
generalizations) compute an encounter until it is unambiguously
over---that is, two ``stable'' objects are receding from one another
with positive velocity at infinity.  A stable object is a star or
merger product, or any binary or dynamically stable multiple whose
components are themselves stable.  Within TRIPLETYCH, the software
automatically detects collisions and close encounters, classifies the
dynamical state of the system, and passes all data to the SH module.
Currently, the dynamical calculation is resumed (via a simple Kepler
solver in the three-body case, or by reverting to the scattering
package in more complex configurations) once dynamical equilibrium is
restored, as determined by MMAS.  Should a second collision occur, the
structure of the new triple merger product is computed similarly.
Once no further interactions are indicated, the SE module is employed
to determine the long-term evolution of the resulting object(s).

The separation of functionality just described is consistent with the
characteristic time scales expected for the dynamical, hydrodynamical,
and stellar-evolutionary processes involved in a simple three-body
scattering.  However, for more complex interactions, it will probably
be desirable to integrate the three modules more closely, for example
using the SE interfaces defined in MODEST-1 and implemented in
TRIP(LE)TYCH by Hurley, including equivalent prescriptions for the
evolution of newly merged systems not yet in thermal equilibrium.

TRIPLETYCH can also be run online via a Web interface.  To
start the simulation the user must choose the parent stars involved,
set the velocity at infinity and impact parameter of the outer orbit,
and set the semi-major axis and eccentricity of the inner orbit.  All
other orbital parameters are chosen at random (but the random seed may
be specified to allow reproducible results).  The Web interface will
be expanded as the description of the underlying physics is refined.

There are still a number of improvements that can be made to these
programs.  It is hoped that web interfaces and free source code will
continue to encourage collaborations as the modules are improved.  The
SD should ultimately be able to handle a true many-body system.  MMAS
should be replaced with a more general SH module that, among other
improvements, allows for the possibility that the two stars do not
merge. The SE code should be replaced with one that uses the full
structure and chemical composition information provided by the SH
module. Furthermore, the SE module will be expanded to allow for
aspects of close binary evolution such as stable mass transfer, tidal
interaction, and gravitational radiation, to name a few (see also
\S4.1).

\section{Science Goals}

MODEST was first conceived to address scientific problems concerning
old globular clusters. It became clear that the MODEST approach was
applicable and relevant to more astrophysical situations than just
globulars, however, including galactic nuclei, young star clusters and
star forming regions. At MODEST-2, these science goals were explored
in more detail. In the following section, observations (mostly of
young objects) with relevance to the MODEST collaboration are
discussed, along with the questions they raise. \S3.2 and 3.3 explore
the questions and some possible solutions to an additional science
goal of the MODEST collaboration -- that is, the specification of
reasonable and realistic initial conditions for models of all dense
stellar systems.
 
\subsection{Observational Motivations}
 
\begin{enumerate}
 
\item The observed high binarity and multiplicity of massive stars
(for visual binaries see Mason et al. 1998, Preibisch et al. 2000; for
spectroscopic binaries in clusters see Garcia \& Mermilliod 2001)
raises the question whether this is due to initial cloud or disk
fragmentation or due to early dynamical evolution (Zinnecker 2002).
In particular, the surprising excess of short-period (5-7 days)
massive double lined spectroscopic binaries in some young clusters
calls for an explanation. Is it due to tidal capture (Zinnecker and
Bate 2002) or due to $N$-body evolution (Bate, Bonnell, Bromm 2002)?

\item In clusters with few massive stars, observations show that
central \astrobj{Trapezium} systems are a common feature (Garcia \&
Mermilliod 2001).  Why is this so, and what is the dynamical evolution
of Trapezium-like configurations? A series of $N$-body models with
different initial conditions may help to answer the last
question. However, the initial configurations can be very complex. For
example, in the \astrobj{Orion Trapezium Cluster} at least one of the
Trapezium members is itself a Trapezium-like subsystem, and the other
members (except \astrobj{$\theta$ Ori 1D}) are binary or triple
systems (see, e.g. Preibisch et al. 2000, Schertl et al. 2003).

\item 
The observed mass segregation in the \astrobj{Orion Trapezium} cluster
(Hillenbrand 1997) and other clusters such as \astrobj{NGC 3603} and
\astrobj{the Arches} (Eisenhauer et al. 1998, Stolte et al. 2002) as
well as the exciting star cluster \astrobj{R136} of the \astrobj{30
Dor} giant HII region (Brandl et al. 1996) raises the question if this
segregation is from birth (`primordial') or due to fast dynamical
evolution. Bonnell \& Davies (1998) did simulations to confirm that
the \astrobj{Trapezium Cluster} does not have the time to evolve dynamically and
that the mass segregation must be primordial, but their Nbody2
calculations should perhaps be repeated with Nbody6 (to check whether
a smaller softening parameter of the gravitational force matters for
mass segregation or not). In addition, a realistic primordial binary
population should be included. The Nbody6 computations of an ONC-like
cluster by Kroupa (2002) suggest that the observed mass segregation
may be obtained dynamically if the embedded cluster is dense enough,
but this issue needs further study.

\item The observed location of the massive \astrobj{IRS16} group of stars as
well as the well-known HeI emission line stars (Allen \& Burton 1994,
Krabbe et al. 1995, Genzel et al. 2000) close to the Galactic Center
is another challenging question: were they formed there or swept into
the inner few parsec region by some sort of disrupted cluster? The
latter is likely, as Portegies Zwart et al. (2002) have shown in their
recent simulation. However, the issue remains as to how close to a
galactic center a massive star cluster can form given the strong tidal
field. For example \astrobj{the Arches} cluster did form only 30
parsec from our Galactic Center, and its tidal radius is only about
2.5 parsec.
 
\item The observed field binary statistics (frequencies, separations, and
mass ratios) of low-mass and intermediate mass stars (F to B) must be
compared with the binary statistics in open clusters (e.g. in
\astrobj{M16} see Duchene et al. 2001) and OB associations (e.g. in
\astrobj{Sco OB2} see Brown 2001) in order to tackle the question of
which mix of progenitor binary populations will provide the correct
field star binary population (this was called 'inverse binary
population synthesis' in \S3.3, see also Ghez 2001; Koehler 2003).  By
understanding the dynamical evolution and dispersal of binary
populations in young clusters and associations, can we retrace the
origin of the field stars in general?
 
\item The observational statistics of runaway OB stars
raises the question of whether isolated OB stars exist or whether they
are the products of dynamical ejection from a nearby star cluster.
Two good examples where B0 stars may have been ejected from young
embedded protoclusters are the \astrobj{S255} and \astrobj{MonR2}
clusters (Zinnecker et al. 1993, Carpenter 2000); and Clarke and
Pringle (1992) suggest that the OB runaways are inconsistent with a
standard IMF in a young open cluster. Are these unusual cases, or the
norm?

\item A large number of millisecond pulsars in globular clusters
(e.g. \astrobj{47 Tuc}, Robinson et al. 1995) have been observed. This
raises the question about what happened in the first fews tens of Myr
in a young globular cluster. Were there several periods of star
formation in globular clusters, i.e extended periods where stars and
gas would coexist?  Which effects would the hybrid evolution of bound
gas and stars suffer (drag, revirialisation of the core, shrinkage,
etc)?

\item The observed null result of radial velocity variations in
34000 stars in \astrobj{47 Tuc} (in an attempt to search for giant
planets, Gilliland et al. 2000) raises the question of the fate of any
planetary mass companions in dense globular clusters.  Have they never
formed or have they all been ejected, creating a population of
free-floating planets in those clusters? See Davies \& Sigurdsson
(2001) and Hurley \& Shara (2002) for relevant simulations.

\end{enumerate} 

To turn the idea of observational motivation around, H. Zinnecker
posed the question: Which kind of HST observations could the MODEST
consortium propose (as a group) to seriously test some of their
results? For example NICMOS observations on embedded protoclusters
should be conducted, such as the one associated with the HII region
\astrobj{G308.70+0.60} at 5kpc, 10 times nearer than \astrobj{30 Doradus}
(Cohen et al. 2002).

The following simulations were also proposed as being of strong
interest to observational questions: 1) simulate dynamical evolution
of toy globular clusters with a truncated IMF (e.g. no stars below 1
$M_{\odot}$, no stars below 0.5 $M_{\odot}$, no intermediate-mass
stars, or some other semi-ridiculous situation) 2) simulate stellar
population synthesis (for galaxies) with and without interacting
binaries (see Portegies Zwart, Yungelson \& Nelemans, 2001).

Though the above issues concentrate on young stellar systems, much of
the work of MODEST is directed to old objects, especially globular
clusters.  One of the classical, observationally motivated problems
here is the construction of dynamical models.  For a long time this
was dominated by King models and its variants, but {\sl evolutionary}
models have been constructed for a number of objects.  One issue here
is how one selects initial conditions that lead, after 12 Gyr of
dynamical and stellar evolution, to models that fit the present
observed structure.

Towards this goal, M. Giersz, E. Vesperini and D. Heggie will be
working on the modelling of specific globular clusters,
i.e. attempting to fit the surface brightness profile, mass functions
and radial velocity dispersion profile.  So far this has been done by
Monte Carlo modelling without binaries (Giersz \& Heggie 2003), and
their intention is to extend this by (a) incorporating binary
populations in the Monte Carlo code (see \S4.1 below) and (b)
cross-checking the results by the slower but less approximate $N$-body
method (\S5).

\subsection{Initial Conditions}

In order to follow the evolution of a dense stellar system, one has to
know (or choose) the initial conditions of that system. Traditionally,
the starting point for a dynamical simulation has been a Plummer model
or a King model. The system includes only stars that are on the
zero-age main sequence and distributed evenly throughout the cluster,
and no gas, star formation, or proto-stellar disks. We know that most
of these assumptions are at best simple and at worst downright
wrong. In this section, we outline the results of star formation and
molecular cloud evolution calculations that are relevant for the
initial conditions of star cluster evolution simulations.

Modern star formation theory considers supersonic interstellar
turbulence, ubiquitously observed in star forming molecular clouds, as
controling agent for stellar birth, rather than mediation by magnetic
fields as was previously assumed, but which fails to predict many of
the observed properties of star forming clouds (Whitworth \etal\ 1996,
Nakano 1998, Crutcher 1999, Bourke \etal\ 2001, Andr{\'e} \etal\ 2000,
Mac~Low \& Klessen 2003).

The key point to this new understanding lies in the properties of
interstellar turbulence that is typically supersonic as well as
super-Alfv{\'e}nic. It is energetic enough to counterbalance gravity
on global scales, but at the same time it may provoke local collapse
on small scales. This apparent paradox can be resolved when
considering that supersonic turbulence establishes a complex network
of interacting shocks, where converging flows generate regions of high
density. This density enhancement can be sufficiently large for
gravitational instability to set in.  The same random flow that
creates density enhancements, however, may disperse them again.  For
local collapse to result in stellar birth, it must progress
sufficiently fast for the region to `decouple' from the flow.  Typical
collapse timescales are therefore of the same order as the lifetimes of
shock-generated density fluctuations in the turbulent gas. This makes
the outcome highly unpredictable. As stars are born through a sequence
of stochastic events, any theory of star formation is in essence a
statistical one with quantitative predictions only possible for an
ensemble of stars.

In this new picture, the efficiency of protostellar core formation,
the growth rates and final masses of the protostars, essentially all
properties of nascent star clusters depend on the intricate interplay
between gravity on the one hand side and the turbulent velocity field
in the cloud on the other. The star formation rate is regulated not
just at the scale of individual star-forming cores through ambipolar
diffusion balancing magnetostatic support, but rather at all scales
(Elmegreen 2002), via the dynamical processes that determine whether
regions of gas become unstable to prompt gravitational collapse. The
presence of magnetic fields does not alter that picture significantly
(Mac~Low \etal\ 1998, Stone \etal\ 1998, Padoan \& Nordlund 1999,
Heitsch \etal\ 2001), as long as they are too weak for magnetostatic
support, which is indicated by observations (Crutcher 1999, Bourke
\etal\ 2001). In particular, magnetic fields cannot prevent the decay
of interstellar turbulence, which in turn needs to be continuously
driven or else stars form quickly and with high efficiency

Inefficient, isolated star formation will occur in regions that are
supported by turbulence carrying most of its energy on very small
scales. This typically requires an unrealistically large number of
driving sources and appears at odds with the measured velocity
structure in molecular clouds which in almost all cases is dominated
by large-scale modes. The dominant pathway to star formation therefore
seems to involve cloud regions large enough to give birth to
aggregates or clusters of stars. This is backed up by careful stellar
population analysis indicating that most stars in the Milky Way formed
in open clusters with a few hundred member stars (Kroupa 1995, Adams
\& Myers 2001).

Clusters of stars build up in molecular cloud regions where
self-gravity overwhelms turbulence, either because such regions are
compressed by a large-scale shock, or because interstellar turbulence
is not replenished and decays on short timescales.  Then, many gas
clumps become gravitationally unstable synchronously and start to
collapse. If the number density is high, collapsing gas clumps may
merge to produce new clumps that now contain multiple protostars.
Mutual dynamical interactions become common, with close encounters
drastically altering the protostellar trajectories, thus changing the
mass accretion rates. This has important consequences for the stellar
mass distribution.  Already in their infancy, i.e.\ in the deeply
embedded phase, very dense stellar clusters are expected to be
strongly influenced by collisional dynamics (Bonnell \etal\ 1997;
Klessen \& Burkert 2000, 2001; Bonnell {\it et al.} 2001a,b; Klessen
2001a,b).

In the following we list some of the recent observational and
theoretical findings that are directly relevant to formation and
evolution of star clusters.

\begin{itemize}
\item Star clusters are expected to build up fast, i.e.\ on timescales
  of order of the free-fall time $\tau_{\rm ff}$ (e.g.\ Klessen \etal\ 
  2000, Bate, Bonnell, \& Bromm 2002), as opposed to the much longer
  ambipolar diffusion timescale proposed by the `standard' theory of
  magnetically mediated star formation (Shu 1977, Shu \etal\ 1987).
  Indeed the observed age spread of stars in young clusters is exactly
  of order $\tau_{\rm ff}$ (for Taurus see Hartmann 2001, 2002,
  however, consider also Palla \& Stahler 1999, 2002; for the
  \astrobj{Trapezium} cluster in Orion see Hillenbrand 1997 or Hillenbrand \&
  Hartmann 1998). For calculations of the subsequent dynamical
  evolution of star clusters this has the important consequence that
  there is a relatively well defined starting time.

\item The pre-main sequence timescale for low-mass stars can reach
  several tens of millions of years. That means that during an initial
  period of a few $\times 10^7\,$years the population of young stellar
  clusters contains both main sequence (MS) as well as pre-main
  sequence (PMS) stars.  As PMS stars in general have
  considerably larger radii than MS stars (e.g.\ Palla 2000), the
  effects of stellar collisions will be strongly enhanced during the
  first few million years of cluster evolution (see e.g.\ the
  discussion in Bonnell \etal\ 2001a,b). This is not taken into
  account in any of the current star cluster evolutionary
  calculations.

\item It remains quite unclear what terminates stellar birth on scales
  of individual star forming regions.  Three main possibilities exist.
  First, feedback from the stars themselves in the form of ionizing
  radiation and stellar outflows may heat and stir surrounding gas up
  sufficiently to prevent further collapse and accretion.  Second,
  accretion might abate either when all the high density,
  gravitationally unstable gas in the region has been accreted in
  individual stars, or after a more dynamical period of competitive
  accretion, leaving any remaining gas to be dispersed by the
  background turbulent flow.  Third, background flows may sweep
  through, destroying the cloud, perhaps in the same way that it was
  created. Most likely the astrophysical truth lies in some
  combination of all three possibilities.
  
\item Most relevant to the formation of rich clusters is gas expulsion by
  radiation and winds from massive stars. The UV flux from O/B stars
  ionizes gas out beyond the local star forming region.  Ionization
  heats the gas, raising its Jeans mass, and possibly preventing
  further protostellar mass growth or new star formation.  The
  termination of accretion by stellar feedback has been suggested at
  least since the calculations of ionization by Oort \& Spitzer
  (1955).  Whitworth (1979) and Yorke \etal\ (1989) computed the
  destructive effects of individual blister H{\sc ii} regions on
  molecular clouds, while in series of papers, Franco \etal\ (1994),
  Rodriguez-Gaspar \etal\ (1995), and Diaz-Miller \etal\ (1998)
  concluded that indeed the ionization from massive stars may limit
  the overall star forming capacity of molecular clouds to about 5\%.
  Matzner (2002) analytically modeled the effects of ionization on
  molecular clouds, concluding as well that turbulence driven by H{\sc
    ii} regions could support and eventually destroy molecular clouds.
  Focusing on the dynamical evolution of young star clusters subject
  to sudden gas removal, Kroupa, Aarseth, \& Hurley (2001)
  demonstrated the existence of an evolutionary sequence that connects
  massive embedded star clusters with the \astrobj{Orion nebula cluster} and the
  \astrobj{Pleiades}. These models treat cluster gas only in form of a smooth
  and time-varying background potential.  The key question remains,
  however, whether H{\sc ii} region expansion couples efficiently to
  clumpy, inhomogeneous molecular clouds. This can only be addressed
  with combined hydro- and stellar dynamical models (see Geyer \&
  Burkert 2001 for a first attempt).
  
\item The theory of turbulent cloud fragmentation furthermore predicts
  massive stars to form towards the cluster center while lower-mass
  stars will build up more dispersed (e.g.\ Klessen 2001b). Star
  clusters are thus believed to have a considerable degree of mass
  segregation already in their embedded phase. This is in agreement
  with recent finding in very young stellar clusters that often
  exhibit a degree of mass segregation that cannot be explained by
  subsequent dynamical evolution, as e.g.\ observed in \astrobj{NGC 330} in
  the \astrobj{Small Magellanic Cloud} (Sirianni \etal\ 2002).
  
\item Star clusters are also expected to form with a high degree of
  substructure (e.g.\ Klessen \& Burkert 2000). This is indeed
  observed in almost all low-mass star forming regions (for
  \astrobj{Taurus} see e.g., Mizuno \etal\ 1995 or Hartmann 2002; for
  \astrobj{$\rho$ Ophiuchus} see Motte, Andr{\'e}, \& Neri 1998 or
  Bontemps \etal\ 2001) and constitutes an important aspect of their
  further dynamical evolution. In rich clusters, however, the
  relaxation timescales are shorter, and such clusters will thus
  experience a considerable degree of relaxation and erasure of
  substructure already in the embedded phase, i.e.\ before the cluster
  becomes fully visible at optical wavelengths.  Focusing on the
  \astrobj{Trapezium} cluster this issue has been discussed by Scally \& Clarke
  (2002).
  
\item Stars typically form as parts of a binary or higher-order
  multiple system. For the Galactic field stars in the solar
  neighborhood the binary frequency is estimated to be about 50\%
  (Duquennoy \& Mayor 1991), the fraction observed in young star
  clusters typically is comparable to that (e.g.\ in the
  \astrobj{Trapezium}, Prosser \etal\ 1994, Petr \etal\ 1998), but may
  reach values as high as 100\% in dilute regions of low-mass star
  formation (e.g.\ in \astrobj{Taurus}, Leinert \etal\ 1993, Ghez
  \etal\ 1993, K{\"o}hler \& Leinert 1998).  For further references
  consult the review by Mathieu \etal\ (2000). This is expected from
  turbulent fragmentation models, and is consistent with cluster
  evolution calculations having a high fraction of `primordial'
  binaries (e.g.\ Kroupa 1998, or Kroupa, Petr, \& McCaughrean 1999).

\end{itemize}

\subsection{An example of a standard reference star-formation model}

\label{sec:standmod}

In the work of Kroupa and collaborators, a particular set of initial
parameters has emerged as a kind of standard model.  A brief
description of those parameters is presented here. It is useful as
standardised and realistic initial conditions for $N-$body
computations of star clusters.  The standard model is defined by a
minimal set of assumptions based on empirical and theoretical evidence
that describe the outcome of star formation. The model has been
developed in Kroupa (1995 $=$K2) by applying inverse dynamical
population synthesis to find the dominant star-formation events that
produced the Galactic field population, taking as an initial boundary
condition the observed pre-main sequence binary-star properties in
\astrobj{Taurus-Auriga}. It accounts for the properties of short-period binary
systems, but does not incorporate brown dwarfs. In the strict form, it
therefore only applies to late-type stars. This model leads to stellar
populations in good agreement with available observational evidence
for Galactic-field stars and pre-main sequence stars in dense clusters
(K2; Kroupa, Aarseth \& Hurley 2001).

The standard model can be used to search for variations of the IMF or
binary-star properties with star-formation conditions.  If a
population is found which has an abnormal IMF or unusual binary-star
properties, and if dynamical and stellar evolution cannot reproduce
these observations given the standard model, then a very strong case
for a variation of the IMF or binary-star properties has been found.
An example of such an application is provided by Kroupa (2001). 

The standard model assumes: 

\begin{enumerate}

\item \label{p1} All stars are paired randomly from the IMF to form
binary systems with primary mass $m_p$ and companion or secondary mass
$m_s \le m_p$.

\item \label{p2} The distribution of orbital elements (period,
eccentricity and mass ratio) does not depend on the mass of the
primary star, but allowance for eigenevolution (see below) is made. 

\item \label{p3} Stellar masses are not correlated with the
phase-space variables (no initial mass segregation in a cluster).

\end{enumerate}

Assumption~\ref{p1} leads to a flat initial mass-ratio distribution
for late-type primaries, $f_{\rm q}$, (fig.~12 in K2), and is in good
agreement with the flat mass-ratio distribution for $q\equiv m_2/m_1 >
0.2$ derived from observational data of pre-main sequence binaries by
Woitas, Leinert \& K\"ohler (2001). They state that ``these findings
are in line with the assumption that for most multiple systems in
T~associations the components' masses are principally determined by
fragmentation during formation and not by the following accretion
processes''. This in turn is supported by the finding that the mass
function of pre-stellar cores in \astrobj{$\rho$~Oph} already has the
same shape as the Galactic-field IMF, thus indicating that the
fragmentation of a molecular cloud core defines the distribution of
stellar masses (Motte, Andr{\'e} \& Neri 1998; Bontemps et al. 2001;
Matzner \& McKee 2000). By extending the standard model to include
brown dwarfs, the stellar pairing properties are changed by allowing
stars to have brown dwarf companions. The fraction of such systems may
be appreciable but depends on the IMF for brown dwarfs. Likewise,
extension of the standard model to massive stars implies that most
O~stars will have low-mass companions.

Assumption~\ref{p2} is posed given the indistinguishable period
distribution function of Galactic-field G-dwarf, K-dwarf and M-dwarf
binary systems (fig.~7 in K2). The discordant period distributions
between the pre-main sequence binaries and the Galactic-field systems
can be nicely explained by disruption of wide-period binaries in small
embedded clusters containing a few hundred stars. This destruction
process also leads to the observed mass-ratio distribution for G-dwarf
primaries in the Galactic field. The model is also in good agreement
with the observed smaller binary fraction of M~dwarfs than of K~dwarfs
and G~dwarfs.

Assumption~\ref{p3} allows investigation of the important issue
whether massive stars need to form at the centres of their embedded
clusters to explain the observed mass segregation in very young
clusters such as the \astrobj{ONC}. Assumption~\ref{p3} is motivated by
observations that indicate that at least some massive stars appear to
be surrounded by massive disks (e.g. Figueredo et al. 2002) suggesting
growth of the massive star by disk accretion rather than through
coagulation of proto-stars, and by the observations that forming
embedded clusters are typically heavily sub-clustered, with massive
stars forming at various locations (e.g. Motte, Schilke \& Lis
2002). On-going $N-$body work is addressing the issue if dynamical
mass-segregation can account for the observed mass segregation in the
\astrobj{ONC} for example, but the alternative scenario is that coagulation of
forming proto-stars in the densest embedded cluster region with
continued accretion of low-angular momentum material onto the forming
cluster core leads to the build-up of a core of massive stars there
(Bonnell, Bate \& Zinnecker 1998; Klessen 2001b).

The initial distribution functions that are needed to describe a
stellar population are the IMF, the period and eccentricity
distribution functions. The IMF is conveniently (for computational
purposes) taken to be a multi-power-law form,
\begin{equation}
\xi (m) = k\left\{
          \begin{array}{l@{\quad\quad,\quad}l}
   \left({m\over m_{\rm H}}\right)^{-\alpha_0}  &m_l < m \le m_{\rm H}, \\
   \left({m\over m_{\rm H}}\right)^{-\alpha_1}  &m_{\rm H} < m \le m_0, \\
   \left[\left({m_0\over m_{\rm H}}\right)^{-\alpha_1}\right] 
        \left({m\over m_0}\right)^{-\alpha_2} 
        &m_0 < m \le m_1,\\
   \left[\left({m_0\over m_{\rm H}}\right)^{-\alpha_1}
        \left({m_1\over m_0}\right)^{-\alpha_2}\right] 
        \left({m\over m_1}\right)^{-\alpha_3} 
        &m_1 < m \le m_2,\\
   \left[\left({m_0\over m_{\rm H}}\right)^{-\alpha_1}
        \left({m_1\over m_0}\right)^{-\alpha_2}
        \left({m_2\over m_1}\right)^{-\alpha_3}\right] 
        \left({m\over m_2}\right)^{-\alpha_4} 
        &m_2 < m \le m_u,\\
          \end{array}\right.
\label{eq:imf_mult}
\end{equation}
where $k$ contains the desired scaling, and $dN=\xi(m)\,dm$ is the
number of stars in the mass interval $m$ to
$m+dm$. Eq.~\ref{eq:imf_mult} is the general form of a five-part
power-law form, but at present observations only support a three-part
power-law IMF (Kroupa 2002) with $m_l=0.01\,M_\odot, m_{\rm
H}=0.08\,M_\odot, m_0=0.5\,M_\odot$, and $\alpha_2=\alpha_3=\alpha_4$,
\begin{equation}
          \begin{array}{l@{\quad\quad,\quad}l}
\alpha_0 = +0.3\pm0.7   &0.01 \le m/M_\odot < 0.08, \\
\alpha_1 = +1.3\pm0.5   &0.08 \le m/M_\odot < 0.50, \\
\alpha_2 = +2.3\pm0.3   &0.50 \le m/M_\odot. \\
          \end{array}
\label{eq:imf}
\end{equation}
The multi-part power-law form is convenient because it allows an
analytic mass-generation function to be used which leads to very
efficient generation of masses from an ensemble of random
deviates. The multi-part power-law form also has the significant
advantage that various parts of the IMF can be changed without
affecting other parts, such as changing the number of massive stars by
varying $\alpha_4$ without affecting the form of the luminosity
function of low-mass stars. Other functional descriptions of the IMF
are in use (e.g. Chabrier 2001).

A convenient form for the initial period distribution function that
has an analytic period-generation function is derived in K2,
\begin{equation}
f_{\rm P,birth} = 2.5 \, { \left(lP - 1 \right)
  \over 45 + \left(lP - 1 \right)^2},
\label{eq:fp}
\end{equation}

where $f_{\rm P,birth}\,dlP$ is the proportion of binaries among all
systems with periods in the range $lP$ to $lP+dlP$ ($P$ in days), and
$1\le lP\equiv {\rm log}_{10}P$. The usual notation for the binary
proportion is used here, $f_{\rm P}=N_{\rm bin, P}/N_{\rm sys}$, where
$N_{\rm sys}=N_{\rm bin}+N_{\rm sing}$ is the number of systems and
$N_{\rm bin, P}$ is the number of binary systems with periods in the
bin $lP$.  The condition $\int_{lP}f_{\rm P,birth}=1$ (all stars being
born in binaries) gives $P_{\rm max}=10^{8.43}$~d for the maximum
period obtained from the distribution given in equation~\ref{eq:fp}.
$N-$body experiments demonstrate that the observed range of periods
($P\approx 10^{0-9}$~d) must be present as a result of the
star-formation process; encounters in very dense sub-groups cannot
sufficiently widen initially more restricted period distributions and
at the same time lead to the observed fraction of binaries in the
Galactic field (Kroupa \& Burkert 2001).  Observations show that the
eccentricity distribution of Galactic-field binary systems is
approximately thermal, $f_{\rm e} = 2\,e$, and $N-$body calculations
demonstrate that such a distribution must be primordial because
encounters of young binaries in their embedded clusters cannot
thermalize an initially different distribution (K2; Kroupa \& Burkert
2001).

Binary systems in the Galactic field with short periods ($P\lesssim
10^3$~d) do show departures from simple pairing by having a
bell-shaped eccentricity distribution and a mass-ratio distribution
that appears to deviate from random sampling from the IMF. This is
apparent most dramatically in the eccentricity--period diagram that shows an upper eccentricity-envelope for short-period binaries
(Duquennoy \& Mayor 1991).  This indicates that
binary-system--internal processes may have evolved a primordial
distribution. Such processes are envelope--envelope or disk--disk
interactions during youth, shared accretion during youth, rapid tidal
circularisation during youth, and slow tidal circularisation during
the main-sequence phase. These system-internal processes that change
the orbital parameters cannot be expressed with only a few equations
given the extremely complex physics involved, but a simple analytical
description is available through the K2-formulation of {\it
eigenevolution--feeding}. Feeding allows the mass of the secondary to
grow, while eigenevolution allows the eccentricity to circularise and
the period to decrease at small peri-astron distances, and merging to
occur if the semi-major axis of the orbit is smaller than~10 Solar
radii. About 3~per cent of initial binaries merge to form a single
star.  The eigenevolved model-main-sequence eccentricity--period
diagram, and the eccentricity and mass-ratio distributions of
short-period systems, agree well with observational data. In
particular, although the minimum period obtained from eq.~\ref{eq:fp}
is $P=10$~d, eigenevolution leads to the correct number of $P<10$~d
periods.  The resulting IMF of all stars shows slight departures from
the input IMF (eq.~\ref{eq:imf}) as a result of the mass-growth
(feeding) of some secondaries, but the deviations are well within the
IMF uncertainties.

\section{Modelling Goals}

\subsection{Interfaces}

A recurring theme throughout the meeting was the desirability of
multiple versions of different kinds of physics (dynamics, stellar
evolution, binary evolution, etc) that were completely modular, so
that they could be swapped in and out, in different combinations, to
test the robustness of our conclusions.  A workable approach to this
issue is the specification of appropriate interfaces.

MODEST-1 defined a simple but robust interface between dynamical and
(single-star) stellar evolutionary modules (Hut et al.~2003).  The
intent was to construct a ``minimally invasive'' standard means for
dynamical integrators to communicate with stellar evolutionary codes,
without placing any restrictions on the internal language, structure,
or algorithms of either.  Hurley's implementation of this interface
for use within TRIPTYCH and TRIPLETYCH is a promising indicator of the
basic soundness of the approach.  The $N$-body codes {\tt kira} and
{\tt NBODY4} each include a binary evolution algorithm and have
successfully demonstrated that modelling of binary evolution in concert
with stellar dynamics is vital for understanding the nature of stellar
populations of star clusters (Hurley et al. 2001; Portegies Zwart et
al. 2001). However, each algorithm is drawn from a particular (and
different) binary population synthesis code, i.e. the approach to this
point has been distinctly non-modular, and computational constraints
have limited the $N$-body method to small-$N$ so far. Full proof of
concept will be realized when the interface is incorporated into the
$N$-body codes and the equivalent Monte-Carlo schemes, in principle
allowing stellar evolutionary algorithms to be exchanged between
radically different dynamical integrators.

One goal that came out of MODEST-2 concerned the variety of binary
evolution packages that exist. We would like to be able to include
binary evolution into any dynamics code that exists, be it $N$-body,
gas, Monte Carlo or whatever. Therefore, there was a call for a
standardized interface between binary evolution and dynamics
calculations, along the lines of the standardized single star
evolution interface developed after MODEST-1.  Although the detailed
information that may be needed is more complex and the range of
possible evolutionary states is much broader, we believe that a simple
interface similar in spirit to that already developed for stellar
evolution is feasible.  

As an illustration of how this can be done, after the meeting
S. Portegies Zwart and D. Heggie constructed an example showing how
the binary evolution packages in starlab can be integrated with some
other code.  For this purpose they constructed a simple three-body
scattering package, based on the scattering cross sections used by
Giersz (1998, 2001) in his Monte Carlo code, and added the binary
evolution routines from starlab.  The resulting code, called McScatter
(for Monte Carlo scatter) has been made available on the MODEST web
site, and further developments are planned.

While the simple code that is being devised by Portegies Zwart and
Heggie is intended for illustrative purposes only, a more elaborate
code of this kind already exists.  In a recent project at
Northwestern University, N. Ivanova, K. Belczynski, V.  Kalogera
and F. Rasio start with a sophisticated population synthesis
code (which can calculate accurately the evolution of a large
population of non-interacting single and binary stars) and add to it a
simplified treatment of dynamical interactions between stars and
binaries in a dense cluster environment. In the Northwestern project,
all relevant interactions (collisions, binary-single and
binary-binary) are implemented in a Monte Carlo fashion
and with simple recipes for determining the outcomes. The cluster is
modeled as a static background and all interactions are assumed to
take place in a core of fixed size and density.  This approach to
study the evolution of the stellar population in a dense cluster core
has two main advantages: (1) it is very fast (the computational time
is spent almost entirely for the evolutionary calculations; the
evolution of $10^5$ binaries for a Hubble time can be calculated in
about 2 days on a single 2Ghz Pentium IV processor); (2) the
dependence of the resulting stellar population on the dynamics and
cluster parameters can be studied easily and systematically, e.g., by
turning on or off one dynamical effect at a time. Among the many
planned applications of this approach is a new study of the formation
and evolution of low-mass X-ray binaries and millisecond pulsars in
globular clusters. The population synthesis code that this project
used as a starting point is the {\tt StarTrack} code developed by
K. Belczynski and V. Kalogera (Belczynski et al. 2002). This code
evolves binaries using standard prescriptions for population synthesis
studies with improved detailed treatments of many important processes
affecting the stellar evolution and binary orbits: common envelope
evolution (based on an $\alpha_{\rm CE}\lambda$-type prescription) and
complete binary mergers; detailed treatment of stable and unstable,
conservative and non-conservative mass transfer phases, thermal
timescale mass transfer; tidal dissipation, synchronization and
circularization; mass and angular momentum loss through stellar winds;
angular momentum loss through gravitation radiation and magnetic
breaking; hyper-critical accretion onto compact objects, asymmetric
core-collapse events, SN explosions and kicks.

\subsection{Primordial Triple and Multiple Systems}

The incidence of triple and higher-multiple systems in the solar
neighborhood is by no means neglible: probably between 5\% and 15\%
of systems are at least triple. A cross-referencing of the
catalogue of 612 multiple stars by Tokovinin (1997) with the Bright Star
Catalogue (BSC; Hoffleit \& Jaschek 1983; the 9110 brightest
stars, more or less) gives 395 entries in common. This shows clearly 
how incomplete the data must be, and suggests that 5\% is very much a 
lower bound. Much smaller but more thoroughly studied samples suggest 
that 10\% is reasonable, but with considerable uncertainty.

It is not clear to what extent these should be considered
`primordial'. Some might be produced in dense star-forming regions
(SFRs) by binary-binary dynamical interactions, but dynamical
evolution of clusters containing even a high proportion of primordial
binaries do not generally produce as many triples as are observed
(e.g. Kroupa 1995). Direct observation of SFRs suggests that triples
are even more common in them than in the field. Consequently it seems
likely that on balance triples are destroyed rather than created in
dynamical encounters. It seems reasonable therefore that until a
really detailed understanding of star formation can give the observed
frequency of binaries {\it and} triples, we should start dynamical
calculations with a distribution of primordial triples as well as
binaries.

Most of the triples in the field, however, are wide systems where the
outer orbits are of size $\gtrsim 100$AU, and should be relatively
quickly destroyed in dense stellar environments.  However, a
proportion have {\it outer} orbits of $\lesssim 10$AU, and these may
be hard enough to survive for some time, and to influence both
dynamical and stellar evolution in dense clusters. A provisional
estimate is that 1 -- 2\% of systems in the solar neighborhood have
outer orbits $\lesssim 10$AU. The proportion seems to be larger among
systems of higher mass (OBA) than lower mass (FGK). We might note that
among the $\sim 50$ O stars brighter than 6th magnitude,
\astrobj{$\tau$ CMa} is a triple (van Leeuwen \& van Genderen 1997)
with an {\it outer} period of only $155 $d; actually, the system is
quadruple, with a fourth body at a few hundred AU. Among similarly
bright B stars \astrobj{$\lambda$ Tau} has an outer period of only $33
$d (Fekel \& Tomkin 1982).  We can list about 50 of the BSC triple
stars in which the outer period is less than $\sim 10$yr, and the
census is by no means complete since third components in orbits of $1
- 10$yr are usually quite hard to recognise. The detection rate of
such triples appears to be currently of order one per year: a recent
bright addition is \astrobj{$\delta$ Lib} (Worek 2001), a classic
Algol that turns out to have a third body in an orbit of $\sim 1000$d.

Triples are likely to be important both for dynamical and for
stellar-evolutionary reasons. Dynamically, this is because they are
usually of higher mass and so are more likely to sink to the centre.
The triple \astrobj{HD109648} (Jha et al. 2000) consists of three F
stars of very similar masses, with periods of 5.5 and $120$d. Such a
system in a moderately old dense cluster should have an important
effect on the dynamical evolution of the cluster. Evolutionarily, the
same system could be important as a potential blue straggler with as
much as three times the turn-off mass. But there are several other
evolutionary channels that are open to such triples, but not to
binaries.

\subsection{Lusus natur\ae}

In this section we discuss a few special cases for which we have
little understanding and for which no obvious modelling technique
currently exists. The main reason to add this section is to prompt new
research. Most special cases in MODEST originate either on the
interface between two well-developed techniques or due to the effect
one part of the model has on the other. These may lead to unexplored
areas of physics or to monstrosities ({\em lusus natur\ae}). We
therefore do not intend to discuss uncertainties in the various
modelling techniques, such as the mixing length in stellar evolution,
the common envelope parameter in binary evolution or the energy
generation of shocks in hydrodynamical calculation.

We have encountered so many bizarre situations in current models that
we cannot list them all in this section; nor can we anticipate on all
possible processes and creation for which no ready continuation of the
model calculations exists. Instead, we will illustrate the {\em lusus
natur\ae} with a few interesting cases.

The most obvious interface problem comes from the improvement in
stellar physics, from single stellar evolution to binary
evolution. Many publications have been written about the {\em zoo}
which originates when two stars are evolved synchronously while taking
variations in the orbital parameters into account. The introduction of
stellar dynamics to binary evolution leads to all kinds of extra
interface problems and to an enormous enlargement of the possibility
of non-standard cases. Some of the most obvious curiosities when
stellar evolution, binary evolution and stellar dynamics are combined
are binaries with two blue stragglers, a blue straggler more than
twice the turn-off mass or two close white dwarf binaries in eccentric
orbits. These cases are rather rare, and in general we are quite well
equipped to handle such situations.

The real {\em lusus natur\ae} are these cases where no ready
methodology is available. An example of this is mass transfer in
binaries which are strongly perturbed by a third star. Such binaries
can easily pick up some eccentricity in the interaction, which then
can affect the characteristics of mass transfer quite dramatically.
There is very little theoretical understanding of the mass transfer in
eccentric binaries, in part because we have no clear examples in the
solar neighborhood which we can study. For this reason also, these
cases do not always attract the attention they require.

On the interface between gas dynamics and stellar dynamics are several
instances for which there is currently no methodology available. What
happens, for example, to the mass liberated in the low velocity
stellar wind of a low-mass star on asymptotic giant branch?  Generally
it is assumed that this gas is blown out of the star cluster and that
its effect on the stellar motions is negligible. However, in galactic
nuclei, for example, this residual gas can strongly affect the model
calculations. It may even change the surface abundances of other stars
in the cluster, as suggested by D'Antona et al. (2002). The main era
in a cluster's life when gas is important is during its formation, as
discussed in detail in \S3.2 and \S3.3. It may be, however, that gas
needs to be considered to some degree throughout the lifetime of the
cluster.

The interface with hydrodynamics and other modes also poses many
opportunities for {\em lusus natur\ae}. Collisions between many
stellar spectral types have been carried out, even between compact
objects. And in some case the collision products are even further
evolved with stellar evolution models. In these models a clear problem
is the enormous amount of angular momentum which the merger product
has to lose in order to become a relatively normal star again (see
\S4.4).  In recent dynamical models it has become clear that runaway
collisions can be quite common. The evolution of a single collision
product is already quite uncertain, let alone a star which has
experienced more than one collision. It is unclear what kinds of
supernovae these runaway products will produce, or if they will be
substantially unusual in any way.  Finally, there are still some
collisions we have not modelled in detail. Particularly, what happens
when a newly-formed neutron star receives a kick from its supernova,
and then immediately runs into a nearby companion? The canonical
understanding is that it will become a Thorne-Zytkow object, but what
does that look like? Is it something we can detect as strange?

As a last case, we mention the interaction between a star cluster and
its direct gravitational environment, such as the tidal field of the
Galaxy, other nearby star clusters or simply the swarm of field stars
in the clusters' surrounding. The first case has been studied in some
detail, but the others require more thought, particularly for studies
of young star clusters near the galactic centre.

It is the goal of the MODEST collaboration to categorize and address
these issues, and to develop the necessary tools to deal with these
{\em lusus natur\ae}. In this section, we have given a flavour of some
of the issues that are yet to be addressed. We expect the list will
continue to grow as the interfaces between stellar evolution, stellar
dynamics and hydrodynamics become more fully entangled.

\subsection{Stellar Evolution of Non-Standard Stars}

One of the goals of the MODEST collaboration is to be able to evolve
stellar collision products and binary merger products on-the-fly when
they are created during the dynamical evolution of a stellar
system. The biggest problem with evolving these products is that they
`begin' their lives significantly out of thermal equilibrium, even if
they are in hydrostatic equilibrium. For a description of evolution
calculations of products of collisions between two main sequence stars
(i.e. blue stragglers), see \cite{SLBDRS97} and
\cite{SFLRW01}. They use the results of SPH simulations of
collisions directly as starting models for stellar evolution
calculations, and follow the collision product through the thermal
relaxation phase to the main sequence and beyond. The results for
head-on collisions are reasonable and robust. When the collisions are
not head-on, however, the collision product has a significant amount
of angular momentum from the initial orbit of the two parent
stars. Since the `proto-blue-straggler' does not have a surface
convection zone, there is no obvious way for it to lose angular
momentum (through a magnetic wind, for example). It needs to lose most
of its angular momentum so that it does not reach break-up velocity as
it contracts to the main sequence. A possible solution to this problem
is to have the proto-blue-straggler create and retain a disk of
material for a few Myr (probably the first material that is thrown
off by the contraction of the rapidly rotating product). If the blue
straggler can become locked to the disk during its contraction phase,
or even a portion of it, the star will spin down by transfering
angular momentum to the disk, in the same manner as protostars
\citep{SPT00, BSP01}. Preliminary calculations of this process are 
giving promising results.

When studying blue stragglers, it is also necessary to consider the
blue stragglers that are formed from the primordial binaries (either
initially close binaries, or ones that have undergone an exchange
during a close encounter, as M. Davies discussed at MODEST-2). The
structure, and hence subsequent evolution, of mass transfer remnants
remains uncertain. Simulations of mass transfer and common envelope
evolution are called for, so that the structure of the products can be
determined accurately.

There are more stars in globular clusters than just main sequence
stars. Giant branch stars, white dwarfs, even neutron stars are
involved in collisions in the dense regions of clusters with
significant regularity. Collisions involving giants, in particular,
may explain some observations of globular clusters. The cores of dense
globular clusters seem to be lacking in bright giants \citep{B94};
some core collapse clusters show evidence for colour gradients, in the
sense of being bluer in the centre \citep{DPPC91}; and extreme
horizontal branch stars (or sdB stars) seem to be concentrated towards
the centres of dense clusters \citep{FFB92}. The suggestion is that
giants are involved in collisions in the densest regions of
clusters. The collision removes some mass from the giant, prohibiting
its ascent to the tip of the giant branch, and producing a low-mass
(i.e. blue) horizontal branch star rather than a regular one. SPH
simulations of collisions with giants, particularly those collisions
that are mediated by a binary system, show that significant amounts of
mass can be removed from the giant. Subsequent evolutionary
calculations of both the stripped giant and the incoming star which
removes the mass will constrain this scenario.

Detailed stellar evolution calculations of collision products, using
hydrodynamics simulations to provide the starting conditions, are very
useful for providing the basis for recipes of stellar collision
product evolution, and for determining the best way for live codes to
handle unusual configurations, particularly those out of thermal
equilibrium. By creating and using detailed models, we can have more
confidence in the results of the cluster evolution simulations. 

\section {Comparison and Validation}

The evolution of dense stellar systems is such a complicated problem
that no exact solutions and few exact constraints are known. Therefore
the reliability of simulations can best be checked by
cross-validation. For this purpose we should aim to devise a small
suite of well specified test problems, and to make available standard
sets of results. These can be used to check that a new code is working
correctly, or that approximate methods give results consistent with
more elaborate methods.
                                                                               
Here we summarise the kinds of problems according to the ingredients
that they can be used to check.  We concentrate on studies which have
resulted in tabular data, as without these the necessary comparisons
tend to be rather qualitative.  (For example, the evolution of an
isolated Plummer model with stars of equal mass has been studied many
times, but results are given usually in graphical form.)  We also
restrict attention to problems that have already been studied by more
than one method or code.

\subsection{ Pure stellar dynamics - single stars }

Though small ($N = 25$) by current $N$-body standards, the experiment
reported by Lecar (1968) was the first example of a collaborative
study, and made plain the chaotic evolution of the system.

The name ``first collaborative experiment" (Heggie et al. 1998, Heggie
2003) is usually applied nowadays to a much later problem devised for
the IAU General Assembly in Kyoto in 1997.  This experiment specified
a reasonably rich ($N\simeq2.5\times10^5$) system of unequal masses in
a tidal field.  This is too large for $N$-body models, which had to be
scaled. This led to the interesting discovery that the dissolution
time does not vary in proportion to the relaxation time, as is usually
assumed, but varies more slowly with $N$ (Baumgardt 2001).  Results
are available on the web ({\tt
http://www.maths.ed.ac.uk/\~{}douglas/experiment.html}).

\subsection{ Pure stellar dynamics - binary and single stars}

One of the Fokker-Planck models studied by Gao et al. (1991) has become
a test case that has been used by Giersz \& Spurzem (2000) to compare
with results from a different (but still approximate) method.
$N$-body results would be useful, but since $N\sim3\times10^5$, this
is not feasible at present.  Giersz \& Spurzem have also conducted
comparisons with the $N$-body models of Heggie \& Aarseth (1992), but
these pre-GRAPE models are much too {\sl small} to be useful nowadays.
                                                                              
There is a need for standardised $N$-body models in this area.  The
second collaborative experiment (Kyoto II, see below) assumes
evolution of single and binary stars, but some partial calculations
using stellar dynamics alone have been completed, all with $N$-body
models, and it is hoped that this ``sub-Kyoto II" may meet this need.
                        
\subsection{ Stellar dynamics and stellar evolution - single stars }

A well specified set of models was formulated and studied by Chernoff
\& Weinberg (1990).  This specification then became the basis of
subsequent $N$-body studies by Aarseth \& Heggie (1998), who used
scaling with $N$.  Other studies with this and other methods are
presented by Takahashi \& Portegies Zwart (2000), Giersz (2001) and
Joshi, Nave \& Rasio (2001).

\subsection{ Stellar dynamics and stellar evolution - single and binary 
stars. }

This is the domain of the Kyoto II collaborative experiment (Heggie
2003), which is an example of a single well-specified problem that
should be amenable to simulation by a wide variety of codes.  It is a
specification for the initial and boundary conditions of a rich (16k)
object with 25\% binaries.  Even though the initial conditions were
agreed at IAU Symposium 208 in 2001, and even though there was
considerable discussion and virtual unanimity about them at that time,
progress has been much slower than with the first collaborative
experiment.  Indeed no complete calculation has been achieved so far,
though there have been considerable numbers of ``partial" calculations,
i.e. those which ignore some aspect of the problem, such as binary
evolution, or calculations that differ in some other way from the
correct specification.  Some problems have been due to the
specification of the tide as a cutoff rather than a field, though
there were sound reasons for this choice.  Others are due to the
specification of the initial conditions in ``astrophysical" units
rather than $N$-body units.  The main bottleneck, however, is the fact
that so few codes (so far) include binary evolution.  This is one
reason why progress on the interface with binary evolution (\S4.1)
is viewed as being so urgent.
                                                                               
\subsection{Comparisons for the future }

While the above examples concentrate on cluster-like problems, another
stellar dynamics problem of growing importance is the evolution of
galactic nuclei.  There is a need for comparable but more appropriate
initial conditions (perhaps including 1 or 2 black holes).

One of the weaknesses of current modelling is the fact that all codes
used for studying dense stellar systems incorporate the same fitting
formulae for stellar evolution (Hurley, Tout \& Pols 2000).  This is
one aspect that cannot be validated empirically.

There is a considerable need for comparative studies using the
different codes now available for binary star evolution.  These tend,
of course, to be based on similar assumptions, and so consistent
results need not imply that the results are entirely trustworthy, but
it would be interesting to know just how great the differences can be.

It should soon be possible to incorporate ``live" SPH codes into
dynamical models.  To test this aspect, and to compare SPH simulations
with other hydrodynamical codes or approximate methods, a few initial
conditions for stellar collisions should be devised.

At present no code incorporates ``live" stellar evolution.  When this
improves, it will be useful to specify collision products to feed to
the stellar evolution codes (say, one blue straggler progenitor and
one collisionally stripped giant).

Further developments, as they occur, will be added to the web page of WG7.

\section{The Future}

The MODEST collaboration will continue to have bi-yearly meetings,
and the schedule for the next few MODEST meetings were outlined, as
follows:

\begin{itemize}
\item MODEST-3, Melbourne, Australia, 9-11 July 2003, hosted by Rosemary Mardling
\item MODEST-4, Lausanne, Switzerland, 12-14 January 2004, hosted by Georges Meylan
\item MODEST-5, Hamilton, Ontario, Canada, 11-14 August 2004, hosted by Alison Sills
\item MODEST-6, Heidelburg, Germany, 17-19 January 2005, hosted by Rainer Spurzem
\item MODEST-7, Evanston, Illinois, USA, 29-31 August 2005, hosted by Fred Rasio
\end{itemize}

One outcome of MODEST-2 was the creation of eight ``working groups'',
designed to focus the interests of the different members of the
collaboration; and to allow all members of the community to find what
they need more quickly and easily. The working groups, listed below,
all have websites that can be reached from the main MODEST website. 

\begin{tabular}{lll}
Working Group & Contact Person & email address\\
\hline
1. Star Formation & Ralf Klessen & rklessen@aip.de \\
2. Stellar Evolution & Onno Pols & o.r.pols@astro.uu.nl \\
3. Stellar Dynamics & Ranier Spurzem & spurzem@ari.uni-heidelberg.de\\
4. Stellar Collisions & Marc Freitag & freitag@ari.uni-heidelberg.de \\
5. Simulating Observations & Simon Portegies Zwart & spz@science.uva.nl\\
\hspace{0.5cm} of Simulations & & \\
6. Data Structures & Peter Teuben & teuben@astro.umd.edu \\
7. Validation & Douglas Heggie & d.c.heggie@ed.ac.uk \\
8. Literature & Melvyn Davies & mbd@astro.le.ac.uk \\
\end{tabular}

\section{Summary}

The second meeting of the MODEST collaboration, devoted to MOdelling
DEnse STellar systems, was held in December 2002 at the University of
Amsterdam, Holland. This paper provides a summary of the meeting,
including the presentations made, the discussions held, and the
conclusions drawn. There was a clear consensus among the participants
that the MODEST approach is reasonable, useful and clearly necessary
for many problems in stellar astrophysics, and is not
confined simply to globular cluster dynamics, but spans many scales
from star formation in turbulent molecular clouds to the dynamics of
galactic nuclei.

The main improvement to the state of models since MODEST-1 was the
expansion of the `toy model' for interfacing stellar evolution,
stellar dynamics and hydrodynamics into a simple working codes called
TRIPTYCH and TRIPLETYCH. The expansion of the MODEST collaboration
from one based in $N$-body stellar dynamics to one that encompasses
many different dynamical methods (Monte Carlo, gaseous and hybrid) is
an improvement to the collaboration, in that the validity of results
can be tested more easily through comparison of standard cases
simulated by many groups.

The main science goals outlined at MODEST-2 involved understanding the
initial conditions for star cluster formation. The interaction between
stars and gas, primordial mass segregation, and the effects of
pre-main sequence stars all need to be considered. In addition, we
need to follow the observations closely. Observations of massive star
formation or blue stragglers and binaries in the field provide
crucial information for both initial conditions and later evolution
of clusters.

The main theoretical and modelling goals included creating standard
interfaces for the different physics modules (particularly binary
evolution) that are needed for this work, so that we can test the
different versions by swapping different implementations in and out of
the codes. The effects of triple and higher order star systems (both
primordial and dynamically created) is becoming more and more
necessary to understand and include as the dynamics simulations become
more complicated. We outlined a list of ``complicated cases'' in \S4.3 --
these are scenarios that we see in the dynamics simulations for which
we do not have a good theoretical understanding. Similarly, there is a
need for detailed stellar evolution calculations of non-standard stars
(collision products, binary merger products, etc), which can form the
basis of the recipes or on-the-fly stellar evolution calculations.

While we have made significant progress in the short existence of this
collaboration, there is still much work to be done, and we hope that
future workshops in this series will continue the congenial atmosphere
that characterized the first two MODEST workshops. We hope to see you
and continue the discussion at MODEST-3!

{\it Acknowledgments} We would like to thank all the participants of
MODEST-2 for a very interesting and useful workshop. Those
participants are:

\begin{tabular}{ll}
  Christian Boily       &   Thijs Kouwenhoven     \\
  Anthony Brown         &   Pavel Kroupa          \\
  Melvyn Davies         &   James Lombardi        \\
  Stefan Deiters        &   Steven McMillan       \\
  Peter Eggleton        &   Garrelt Mellema       \\
  Marc Freitag          &   Onno Pols             \\
  Mirek Giersz          &   Alison Sills          \\
  Alessia Gualandris    &   Piero Spinnato        \\
  Douglas Heggie        &   Rainer Spurzem        \\
  Edward van den Heuvel &   Peter Teuben          \\
  Jarrod Hurley         &   Simon Portegies Zwart \\
  Piet Hut              &   Tjeerd van Albada     \\
  Natasha Ivanova       &   Enrico Vesperini      \\
  Lex Kaper             &   Ralf Wijer            \\
  Ralf Klessen          &   Hans Zinnecker        \\
\end{tabular}

\end{document}